\documentclass[11pt]{article}

\usepackage[sort,longnamesfirst]{natbib}

\usepackage{amsbsy,amsmath,amsthm,amssymb,graphicx,verbatim,color}
\usepackage{subfig}
\usepackage{multirow}

\usepackage{geometry}
\newtheorem{theorem}{Theorem}[section]

\newtheorem{lemma}{Lemma}[section]

\newtheorem{proposition}{Proposition}[section]

\theoremstyle{remark}
\newtheorem{example}{Example}[section]

\newfont{\msbm}{msbm10 at 12pt}

\begin{document}
\title{Gibbs Sampling for a Bayesian Hierarchical General Linear Model} 

\author{Alicia A. Johnson \\ Department of Mathematics, Statistics, and Computer
  Science \\ Macalester College \\ {\tt ajohns24@macalester.edu} 
  \\
  \\
  Galin L. Jones \\ School of Statistics \\ University of Minnesota \\
  {\tt galin@stat.umn.edu}} 

\date{Draft: \today} 
\maketitle

\abstract{   We consider a Bayesian hierarchical version of the normal theory
  general linear model which is practically relevant in the sense that
  it is general enough to have many applications and it is not
  straightforward to sample directly from the corresponding posterior
  distribution.  Thus we study a block Gibbs sampler that has the
  posterior as its invariant distribution.  In particular, we
  establish that the Gibbs sampler converges at a geometric rate.
  This allows us to establish conditions for a central limit theorem
  for the ergodic averages used to estimate features of the posterior.
  Geometric ergodicity is also a key component for using batch means
  methods to consistently estimate the variance of the asymptotic
  normal distribution. Together, our results give practitioners the
  tools to be as confident in inferences based on the observations
  from the Gibbs sampler as they would be with inferences based on
  random samples from the posterior.  Our theoretical results are
  illustrated with an application to data on the cost of health plans
  issued by health maintenance organizations. }




\section{Introduction}
\label{sec:introduction}

The flexibility of Bayesian hierarchical models makes them widely
applicable. One of the most popular \citep[see,
e.g.,][]{gelm:etal:2004,spieg:etal} is a version of the usual normal
theory general linear model. Let $Y$ denote an $N \times 1$ response
vector and suppose $\beta$ is a $p \times 1$ vector of regression
coefficients, $u$ is a $k \times 1$ vector, $X$ is a known $N \times
p$ design matrix having full column rank, and $Z$ is a known $N \times
k$ matrix.  Then for $r,s,t \in \{1,2,\cdots\}$, the hierarchy is
\begin{equation}\label{eq:hier}
\begin{split}
Y|\beta,u,\lambda_R,\lambda_D  & \sim \text{N}_N\left(X\beta + Zu,
\lambda_R^{-1}I_{N}\right) \\ 
\beta| u,\lambda_R,\lambda_D   & \sim \sum_{i=1}^r \eta_i
\text{N}_p\left(b_i,B^{-1}\right)\\   
u|\lambda_R,\lambda_D          & \sim
\text{N}_k\left(0,\lambda_D^{-1}I_{k}\right) \\ 
\lambda_R     & \sim \sum_{j=1}^s
\phi_j\text{Gamma}\left(r_{j1},r_{j2}\right) \\ 
\lambda_D     & \sim \sum_{l=1}^t
\psi_l\text{Gamma}\left(d_{l1},d_{l2}\right) \\ 
\end{split}
\end{equation}
where the mixture parameters $\eta_i$, $\phi_j$, and $\psi_l$ are
known nonnegative constants which satisfy 
\[
\sum_{i=1}^r \eta_i = \sum_{j=1}^s \phi_j = \sum_{l=1}^t \psi_l=1 
\]  
and we say $W \sim \text{Gamma}(a,b)$ if it has density proportional
to $w^{a-1}e^{-bw}$ for $w > 0$.  Further, we require $\beta$ and $u$
to be a posteriori conditionally independent given $\lambda_R$,
$\lambda_D$, and $y$ which holds if and only if $X^TZ=0$. Finally,
$b_i \in \mathbb{R}$ and positive definite matrix $B$ are known and
the hyperparameters $r_{j1}$, $r_{j2}$, $d_{l1}$, and $d_{l2}$ are all
assumed to be positive.

Let $\xi=\left(u^T,\beta^T\right)^T$ and $\lambda = \left(\lambda_R,
  \lambda_D \right)^T$.  Then the posterior has support $\mathcal{X} =
\mathbb{R}^{k+p} \times \mathbb{R}_+^2$ and a density characterized by
\[
\pi(\xi,\lambda | y) \propto f(y|\xi,\lambda) f(\xi|\lambda)f(\lambda) \; 
\]
where $y$ is the observed data and $f$ denotes a generic density.
Posterior inference is often based on the expectation of a function
$g: \mathcal{X} \to \mathbb{R}$ with respect to the posterior. For the
model \eqref{eq:hier} we can only rarely calculate the expectation
\[
\text{E}_\pi g(\xi,\lambda) : = \int_\mathcal{X} g(\xi,\lambda) \pi
(\xi,\lambda|y) d\xi d\lambda\, ,
\]
since it is a ratio of two potentially high-dimensional intractable
integrals.  Hence inference regarding the posterior may require Markov
chain Monte Carlo (MCMC) methods.  We consider two-component Gibbs
sampling which produces a Harris ergodic Markov chain
$\Phi=\{(\xi_0,\lambda_0),(\xi_1,\lambda_1),\cdots\}$ with invariant
density $\pi(\xi,\lambda|y)$.  

Suppose $E_{\pi} |g| < \infty$ and we obtain $n$ observations from the
Gibbs sampler.  Then a natural estimate of $E_{\pi} g$ is $\bar{g}_{n}
= n^{-1} \sum_{i=0}^{n-1} g(\xi_{i}, \lambda_{i})$ since $\bar{g}_{n}
\to E_{\pi} g$ with probability 1 as $n \to \infty$. In other words,
the longer we run the Gibbs sampler, the better our estimate is likely
to be.  However, this gives no indication of how large $n$ must be to
ensure the Monte Carlo error $\bar{g}_{n} - E_{\pi} g$ is sufficiently
small.  The size of this error is usually judged by appealing to its
approximate sampling distribution via a Markov chain central limit
theorem (CLT), which in the cases of current interest takes the form
\begin{equation}\label{eq:intro clt}
\sqrt{n} (\bar{g}_{n} - E_{\pi}g) \stackrel{d}{\to}
\text{N}(0, \sigma_{g}^{2}) \hspace*{5mm} \text{ as } n \to \infty
\end{equation}
where $\sigma_{g}^{2} \in (0, \infty)$. Due to the serial correlation
in $\Phi$, the variance $\sigma_{g}^{2}$ will be complicated and
require specialized techniques (such as batch means or spectral
methods) to estimate consistently with $\hat{\sigma}^{2}_{n}$, say.
Suppose $\hat{\sigma}^{2}_{n} \to \sigma_{g}^{2}$ with probability 1
as $n \to \infty$.  In this case, an asymptotically valid Monte Carlo
standard error (MCSE) is given by $\hat{\sigma}_{n} / \sqrt{n}$.  In
turn, this can be used to perform statistical analysis of the Monte
Carlo error and to implement rigorous sequential stopping rules for
determining the length of simulation required
\citep[see][]{fleg:hara:jone:2008,jone:hobe:2001} so that the user
will have as much confidence in the simulation results as if the
observations were a random sample from the posterior; this is
described in more detail in Section~\ref{sec:clt}.

Unfortunately, for Harris ergodic Markov chains simple moment
conditions are not sufficient to ensure an asymptotic distribution for
the Monte Carlo error or that we can consistently estimate
$\sigma_{g}^{2}$.  In addition, we need to know that the convergence
of $\Phi$ occurs rapidly.  Thus, one of our goals is to establish
verifiable conditions under which the Gibbs sampler is geometrically
ergodic, that is, it converges to the posterior in total variation
norm at a geometric rate.

We know of three papers that address geometric ergodicity of Gibbs
samplers in the context of the normal theory linear model with proper
priors. These are \cite{hobe:geye:1998}, \cite{jone:hobe:2004}, and
\cite{papa:robe:2008}.  The linear model we consider substantively
differs from those in \cite{papa:robe:2008} in that we do not assume
the variance components are known.  Our model is also much more
general than the one-way random effects model in \cite{hobe:geye:1998}
and \cite{jone:hobe:2004}.  Gibbs sampling for the balanced one-way
random effects model is also considered in \cite{rose:1995b} where
coupling techniques were used to establish upper bounds on the total
variation distance to stationarity.  However, these results fall short
of establishing geometric ergodicity of the associated Markov chain.

The rest of this paper is organized as follows.  Gibbs sampling for
the Bayesian hierarchical general linear model is discussed in Section
\ref{sec:Gibbs} and geometric ergodicity for these Gibbs samplers is
established in Section \ref{sec:DriftAndMinorization}.  Conditions for
the CLT \eqref{eq:intro clt} are given in Section~\ref{sec:clt} along
with a description of the method of batch means for estimating the
variance of the asymptotic normal distribution.  Finally, our results
are illustrated with a numerical example in Section \ref{sec:example}.
Many technical details are deferred to the appendix.

\section{The Gibbs Samplers}
\label{sec:Gibbs}

The full conditional densities required for implementation of the
two-component block Gibbs sampler are as follows: Conditional on $\xi$
and $y$, $\lambda$ follows the distribution corresponding to density  
\begin{equation}\label{eq:FullCond1}
f(\lambda|\xi,y) = \sum_{j=1}^s\sum_{l=1}^t \phi_j\psi_l
f_{1j}(\lambda_R|\xi,y) f_{2l}(\lambda_D|\xi,y) 
\end{equation}
where $f_{1j}(\cdot|\xi,y)$ denotes a Gamma($r_{j1} + N/2, r_{j2} +
v_1(\xi)/2 $) density and $f_{2l}(\cdot|\xi,y)$ denotes a
Gamma($d_{l1} + k/2, d_{l2} + v_2(\xi)/2 $) density with
 \begin{equation}\label{eq:v}
  v_1(\xi) := (y-X\beta-Zu)^T(y-X\beta-Zu), \hspace{10mm}  v_2(\xi) :=
  u^Tu \; .
\end{equation}  
Also, 
\[
\xi|\lambda,y   \sim \sum_{i=1}^r \eta_i
\text{N}_{k+p}(m_{i},\Sigma^{-1}) 
\]
where
\begin{equation}\label{eq:norm1}
\begin{split}
\Sigma^{-1} & = \left( \begin{array}{cc}
\left(\lambda_RZ^TZ + \lambda_DI_k\right)^{-1} & 0 \\
0               & \left(\lambda_RX^TX+B\right)^{-1} \\
\end{array} \right) \\
m_{i} & = \left( \begin{array}{c}
\lambda_R\left(\lambda_RZ^TZ + \lambda_DI_k\right)^{-1}Z^Ty \\
\left(\lambda_RX^TX + B \right)^{-1}(\lambda_R X^Ty + Bb_i) \\ 
\end{array} \right) \; . \\
\end{split}
\end{equation}
These follow from our assumption that $X^TZ=0$. 

There are two possible update orders for our 2-component Gibbs
sampler.  First, let $\Phi_{1}$ denote the Markov chain produced by
the Gibbs sampler which updates $\xi$ followed by $\lambda$ in each
iteration so that a one-step transition looks like $(\xi',\lambda')
\to (\xi,\lambda') \to (\xi, \lambda)$.  Then the one-step Markov
transition density (Mtd) for $\Phi_{1}$ is
\[
k_1(\xi,\lambda|\xi',\lambda')=f(\xi|\lambda',y) f(\lambda|\xi,y) \; .
\]
Similarly, let $\Phi_{2}$ denote the Markov chain produced by the
Gibbs sampler which updates $\lambda$ followed by $\xi$ in each
iteration so that the one-step transition is $(\xi',\lambda') \to
(\xi',\lambda) \to (\xi,\lambda)$.  Then the corresponding Mtd is
\[
k_2(\xi,\lambda|\xi',\lambda')=f(\lambda|\xi',y) f(\xi|\lambda,y) \; . 
\]
Also, let $\Phi_\xi = \{\xi_0,\xi_1,\cdots\}$ and $\Phi_\lambda =
\{\lambda_0,\lambda_1,\cdots\}$ denote the associated marginal chains
with Mtds
\[
k_\xi(\xi|\xi')=\int_{\mathbb{R}_+^2} f(\lambda|\xi',y)
f(\xi|\lambda,y)\, d\lambda 
\]
and
\[
k_\lambda(\lambda|\lambda')=\int_{\mathbb{R}^{k+p}} f(\xi|\lambda',y)
f(\lambda|\xi,y)\, d\xi \; ,
\]
respectively.  

Because the Mtd's are strictly positive on the state space it is
straightforward to show that $\Phi_{1}$ and $\Phi_{2}$ are Harris
ergodic.  The posterior density $\pi(\xi,\lambda| y)$ is invariant for
$\Phi_{1}$ and $\Phi_{2}$ by construction.  Moreover, it is easy to
see that both chains are Feller.  Similarly, $\Phi_\xi$ and
$\Phi_\lambda$ are Harris ergodic and Feller with invariant densities
the marginal posteriors $\pi(\xi|y)$ and $\pi(\lambda|y)$,
respectively.  Hence all four Markov chains converge in total
variation norm to their respective invariant distributions.  In the
next section we establish conditions under which this convergence
occurs at a geometric rate.

\section{Geometric Ergodicity}
\label{sec:DriftAndMinorization}

\subsection{Establishing Geometric Ergodicity}
Our main goal in this section is to establish conditions for the
geometric ergodicity of $\Phi_1$ and $\Phi_2$. Before doing so it is
useful to acquaint ourselves a concept introduced by
\citet{robe:rose:2001}.  Let $X=\{X_{n}, \, n \ge 0\}$ be a Markov
chain on a space $\mathcal{X}$ and $Y=\{Y_{n}, \, n\ge 0\}$ a
stochastic process on a possibly different space $\mathcal{Y}$.  Then
$Y$ is \textit{de-initializing} for $X$ if, for each $n \ge 1$,
conditionally on $Y_{n}$ it follows that $X_{n}$ is independent of
$X_{0}$.  Roughly speaking, \citet{robe:rose:2001} use this concept to
show that $Y$ controls the convergence properties of the Markov chain
$X$.

To establish the geometric ergodicity of $\Phi_1$ and $\Phi_2$ it
suffices to work with the marginal chains $\Phi_\xi$ and
$\Phi_\lambda$.  First, $\Phi_\xi$ is de-initializing for $\Phi_1$ and
$\Phi_\lambda$ is de-initializing for $\Phi_2$. Results in
\citet{robe:rose:2001} imply that if $\Phi_\xi$ ($\Phi_\lambda$) is
geometrically ergodic, so is $\Phi_1$ ($\Phi_2$).  Further, $\Phi_1$
and $\Phi_2$ are co-de-initializing.  Hence if one is geometrically
ergodic, then they both are and Lemma~\ref{lem:roberose} follows
directly.

\begin{lemma}\label{lem:roberose}
  If $\Phi_\xi$ or $\Phi_\lambda$ is geometrically ergodic, then so are
  $\Phi_1$ and $\Phi_2$.
\end{lemma}

Accordingly, we can proceed by studying the convergence behavior of
the marginal chains. We establish geometric ergodicity for $\Phi_\xi$
by establishing a \textit{drift condition}.  That is we need to
specify a function $V: \mathbb{R}^{k+p} \to \mathbb{R}_+$ and 
constants $0 < \gamma < 1$ and $L < \infty$ such that
\begin{equation}\label{eq:drift1}
  \text{E}[V(\xi) \mid \xi'] \le \gamma V(\xi') +   L   \hspace*{4mm}
  \text{for all   } \xi' \in \mathbb{R}^{k+p}  
\end{equation}
where the expectation is taken with respect to the Mtd $k_\xi$.  Let
$W(\xi) = 1 + V(\xi)$, $b=L+1-\gamma$ and $C = \{\xi \, : \, W(\xi)
\le 4b/(1-\gamma)\}$.  \citet[][Lemma 3.1]{jone:hobe:2004} show that
equation \eqref{eq:drift1} implies
\[
\Delta W(\xi') := \text{E}[W(\xi) \mid \xi'] - W(\xi') \le -
\frac{1-\gamma}{2} W(\xi') + 2 b I(\xi' \in C) \; .
\]
Here $\Delta W(\xi')$ is the \textit{drift}, $V$ (or $W$) is a
\textit{drift function} and $\gamma$ a \textit{drift rate}. If $\xi'
\notin C$ the expected change in $W$ is negative so $\Phi_{\xi}$ will
tend to ``drift'' to $C$, that is, where the value of $W$ is small.
Moreover, it also does it in such a way that the drift towards $C$ is
faster when $\gamma$ is small.  On the other hand, if $\gamma \approx
1$ the drift will be slow.  Thus the value of $\gamma$ is intimately
connected to the convergence rate of $\Phi_{\xi}$; for a thorough
accessible discussion of the connection see \citet[][Section
3.3]{jone:hobe:2001}. Hence examination of $\gamma$ can give us some
intuition for the convergence behavior of $\Phi_{\xi}$.  However, drift
functions are not unique so this examination generally will not lead
to definitive conclusions.

A function $V: \mathbb{R}^{k+p} \to \mathbb{R}$ is \textit{unbounded
  off compact sets} if the set $\{\xi \in \mathbb{R}^{k+p}: V(\xi) \le
d\}$ is compact for any $d>0$. Note that the maximal irreducibility
measure for $\Phi_{\xi}$ is equivalent to Lebesgue on
$\mathbb{R}^{k+p}$ so that its support certainly has a non-empty
interior. The sufficiency of drift for geometric ergodicity now
follows easily from Lemma 15.2.8 and Theorems 6.0.1 and 15.0.1 of
\citet{meyn:twee:1993} and Lemma~\ref{lem:roberose}.

\begin{proposition}\label{prop:geoerg} 
  Suppose \eqref{eq:drift1} holds for a drift function that is
  unbounded off compact sets.  Then $\Phi_\xi$ is geometrically
  ergodic and so are $\Phi_{1}$ and $\Phi_{2}$.
\end{proposition}

In Section \ref{sec:xi} we develop conditions on our Bayesian model
\eqref{eq:hier} which are sufficient for the conditions of
Proposition~\ref{prop:geoerg}.

\subsection{Drift for $\Phi_\xi$}\label{sec:xi}
  
For all $j \in \{1,\cdots,s\}$ and $l \in \{1,\cdots,t\}$, define
constants 
\begin{equation}
 \label{eq:delts}
\begin{split}
  \delta_{j1} & =\frac{\sum_{i=1}^N
    z_i\left(Z^TZ\right)^{-1}z_i^T}{2r_{j1}+N-2};  \hspace*{14mm}
  \delta_{l2}  = \frac{k}{2d_{l1}+k-2}; \\
  \delta_{j3} & =  \frac{\sum_{i=1}^N
    x_i\left(X^TX\right)^{-1}x_i^T}{4(2r_{j1}+N-2)};  \; \text{ and }
  \hspace*{2mm}
  \delta_{l4}  = \frac{k + \sum_{i=1}^N z_iz_i^T}{2d_{l1}+k-2} \; .\\
 \end{split}
\end{equation}
Also, let $x_i$ and $z_i$ denote the $i$th rows of matrices $X$ and
$Z$, respectively, and let $y_i$ and $u_i$ denote the $i$th elements
of vectors $y$ and $u$, respectively.  Next, for $i \in
\{1,\cdots,r\}$ define
\[
G_{i} (\lambda) := \sum_{m=1}^N\left[
  \text{E}_i\;(y_m-x_m\beta-z_mu|\lambda,y)\right]^2 +
\sum_{m=1}^k\left[ \text{E}_i\;(u_m|\lambda,y)\right]^2
\]
where $\text{E}_i$ denotes expectation with respect to the
$N_{k+p}(m_{i},\Sigma^{-1})$ distribution.

\begin{proposition}\label{prop:drift1}
  Assume there exists some $K < \infty$ such that $G_{i}
  (\lambda) \le K$ for all $\lambda \in \mathbb{R}_+^2$ and
  $i \in \{1,\cdots,r\}$.
Let $V(\xi) = v_1(\xi) + v_2(\xi)$ where $v_1(\cdot)$ and $v_2(\cdot)$ are
defined at \eqref{eq:v}.  
\begin{enumerate}
\item If $Z^TZ$ is nonsingular,  $d_{l1}> 1$ for all $l \in
  \{1,\cdots,t\}$, and
\[
r_{j1} > 0 \vee 0.5\left(\sum_{i=1}^N z_i (Z^TZ)^{-1} z_i^T -N+2
\right) \hspace{4mm} \text{ for all } j\in\{1,\cdots,s\} \, ,
\]
 then \eqref{eq:drift1} holds for drift function $V(\xi)$ with
$\max_{j,l}\{\delta_{j1}, \delta_{l2}\} \le \gamma < 1$ and  
\[
 L=\sum_{i=1}^N  x_iB^{-1}x_i^T + \max_{j,l}\left\lbrace
   2r_{j2}\delta_{j1} + 2d_{l2}\delta_{l2}\right\rbrace + K  \;
 . 
\] 
\item If for all $j \in \{1,\cdots,s\}$ and $l\in\{1,\cdots,t\}$
\[
\begin{split}
 r_{j1}  & > 0 \, \vee \,0.5\left[0.25\sum_{i=1}^N x_i (X^TX)^{-1} x_i^T
   -N+2 \right] \;\; \text{and} \\
 d_{l1}  & > 0.5\left[ 2 + \sum_{i=1}^N z_iz_i^T \right] \; . \\
\end{split}
\]
then \eqref{eq:drift1} holds for drift function $V(\xi)$ with
$\max_{j,l}\{\delta_{j3}, \delta_{l4}\} \le \gamma < 1$ and 
\[
L= \frac{1}{4} \sum_{i=1}^Nx_iB^{-1}x_i^T + \max_{j,l}\left\lbrace
  2r_{j2}\delta_{j3} + 2d_{l2}\delta_{l4} \right\rbrace+ K \;
. 
\]
\end{enumerate}
\end{proposition}

\begin{proof} See Appendix~\ref{app:bound}.\end{proof}

Notice that the formulations of $\gamma$ given by Proposition
\ref{prop:drift1} depend on the Bayesian model setting through
$\delta_{j1}$, $\delta_{l2}$, $\delta_{j3}$, and $\delta_{l4}$.
Therefore, the drift and convergence rates of the $\Phi_\xi$ marginal
chain (hence the Gibbs samplers) may be sensitive to changes in the
dimension $k$ of $u$, the total number of observations $N$, or the
hyperparameter setting.  However, it is interesting that the dimension
of $\beta$, which is $p$, has only an indirect impact on this result.
Specifically, when $Z^{T}Z$ is nonsingular the value of $p$ has no
impact, that is, the drift rate is unaffected by changes in $p$.  Of
course, changing $p$ does mean that $X$ changes which may impact
$\delta_{j3}$ which in turn can change the permissible hyperparameters
$r_{j1}$ and the drift rate when $Z^{T}Z$ is singular.

\begin{example}
\label{ex:bri}
Consider the balanced random intercept model derived from
\eqref{eq:hier} for $k$ subjects with $m$ observations each.  In this
case, $Z = I_k \otimes 1_m$ where $\otimes$ denotes the Kronecker
product and $1_m$ represents a vector of ones of length $m$.  Hence
$Z^TZ = mI_k$ is nonsingular.  Define
\[
M_{N,k}:= \max_{l} \left\lbrace \frac{k}{2 r_{j1} + N - 2}, \;
  \frac{k}{2 d_{l1} + k - 2}\right\rbrace \; .
\]
If $d_{l1}>1$ for all $l$, Condition 1 of Proposition
\ref{prop:drift1} establishes drift rate $M_{N,k} \le \gamma <
1$. Notice that $M_{N,k} \to 1$ as $k \to \infty$ and hence $\gamma
\to 1$ as well.  This supports our intuition that the Gibbs sampler
should converge more slowly as its dimension increases.  On the other
hand, if $k$ is held constant but $m$ increases so that $N=km \to
\infty$, then
\[
\frac{k}{2 d_{l1} + k - 2} \le \gamma < 1 \; .
\]
Thus increasing the number of observations per subject does not have
the same negative, qualitative impact as increasing the number of
subjects.  Finally, $M_{N,k} \to 1$ (hence $\gamma \to 1$) when $k$ is
held constant and $d_{l1} \to 1$ for any $l$.
\end{example}

Consider the condition that $G_{i} (\lambda) \le K$ for all $\lambda \in \mathbb{R}_+^2$ and $i \in \{1,\cdots,r\}$. The following result establishes this condition for an important special case of \eqref{eq:hier}.  In our experience it is often straightforward to show that $G_i$ is bounded and, if desired, numerical optimization methods yields appropriate $K$.

\begin{proposition}\label{prop:bound1}
Assume $b_i=0$ for all $i \in \{1,\cdots,r\}$. 
\begin{enumerate} 
\item If $Z=0$, then 
\[
G_{i}(\lambda_{R}) \le y^{T}y 
\]
for all $\lambda_{R} \in \mathbb{R}_+$ and $i \in \{1,\cdots,r\}$.
\item If $Z^TZ$ is nonsingular, then
\[
G_{i} (\lambda) \le  y^Ty + y^TZ(Z^TZ)^{-2}Z^Ty \
\]
for all $\lambda \in \mathbb{R}_+^2$ and $i \in \{1,\cdots,r\}$.
\end{enumerate}
\end{proposition}

\begin{proof} See Appendix~\ref{app:bound}. \end{proof}

We are now in position to state conditions on \eqref{eq:hier}
guaranteeing geometric ergodicity of the Gibbs samplers $\Phi_{1}$ and
$\Phi_{2}$.  This follows easily from Propositions \ref{prop:geoerg}
and \ref{prop:drift1} if the drift function $V(\xi)=
v_1(\xi)+v_2(\xi)$ is unbounded off compact sets on
$\mathbb{R}^{k+p}$. Define $S=\{ \xi \in \mathbb{R}^{k+p} \, : \,
V(\xi) = v_{1}(\xi) + v_{2}(\xi) \le d \}$ where $d>0$. Notice that
$V$ is continuous so it is sufficient to show that, on $S$,
$|\beta_{i}|$ is bounded for $i \in \{1,2,\ldots,p\}$ and $|u_{j}|$ is
bounded for $j \in \{1,2,\ldots,k\}$.  Clearly, $S \subset S_{2} = \{
\xi \, : \, u^{T}u \le d\}$ and it is obvious that each $|u_{j}|$ is
bounded on $S_{2}$ hence also on $S$.  Moreover, note $v_{2} \to
\infty$ as $|u_{j}| \to \infty$.  Given that the $|u_{j}|$ are bounded
it is easy to see that $v_{1} \to \infty$ as $|\beta_{i}| \to
\infty$. Putting this together we see that $V$ is unbounded off
compact sets.  The main result of this section follows.

\begin{theorem}
\label{thm:geo1}
Assume the conditions of Proposition~\ref{prop:geoerg}.  Then the Markov chain $\Phi_{\xi}$ and the Gibbs samplers $\Phi_1$ and $\Phi_2$ are geometrically ergodic.
\end{theorem}

\section{Interval Estimation} \label{sec:clt}

Suppose we want to estimate an expectation $E_{\pi} g : =
\int_{\mathcal{X}} g(\xi, \lambda) \pi(\xi, \lambda | y) d \xi
d\lambda$ where $g$ is real-valued and $\pi$-integrable.  It is
straightforward to estimate $E_{\pi} g$ with $\bar{g}_{n} : = n^{-1}
\sum_{i=0}^{n-1} g(\xi_{i}, \lambda_{i})$.  A key step in the
statistical analysis of $\bar{g}_{n}$ is the assessment of the Monte
Carlo error $\bar{g}_{n} - E_{\pi}g$ through its approximate sampling
distribution. 

\begin{theorem} \label{thm:clt}
Assume the conditions of Theorem~\ref{thm:geo1}.  If $E_{\pi}
|g|^{2+\epsilon} < \infty$ for some $\epsilon > 0$, then there is a
constant $\sigma_{g}^{2} \in (0,\infty)$ such that for any
initial distribution
\[
\sqrt{n}(\bar{g}_{n} - E_{\pi} g) \stackrel{d}{\to} \text{N}(0,
\sigma_{g}^{2}) \hspace{3mm} \text{ as } n \to \infty \; .
\]
\end{theorem}

The proof of this theorem follows easily from Theorem~\ref{thm:geo1},
Theorem 2 of \citet{chan:geye:1994} and Section 1 of
\citet{fleg:jone:2009}.  Roughly speaking, results in
\citet{hobe:etal:2002}, \citet{jone:hara:caff:neat:2006} and
\citet{bedn:latu:2007} show that, under conditions comparable to
those required for Theorem~\ref{thm:clt}, techniques such as
regenerative simulation and batch means can be used to construct an
estimator of $\sigma_{g}^{2}$, say $\hat{\sigma}^{2}_{n}$, such that
$\hat{\sigma}^{2}_{n} \to \sigma_{g}^{2}$ as $n \to \infty$ almost
surely.  See \citet{fleg:jone:2009} for the conditions required to
ensure consistency of overlapping batch means and spectral estimators
of $\sigma_{g}^{2}$.  

Before giving a precise discussion of the conditions for consistency
we need a preliminary definition and result.  Let $\mathcal{X}
\subseteq \mathbb{R}^{d}$ for $d \ge 1$ and $k : \mathcal{X} \times
\mathcal{X} \to [0,\infty)$ be an Mtd with respect to Lebesgue
measure.  Suppose there exists a function $s \, : \, \mathcal{X} \to [0,1)$
and a density $q$ such that for all $x, x' \in \mathcal{X}$
\[
k(x|x') \ge s(x') q(x) \; .
\] 
Then we say there is a \textit{minorization condition} for $k$.

\begin{lemma} \label{lem:silly minorization} 
Let $C \subseteq \mathcal{X}$ be compact and assume $c > 0$ where
\[
c = \int_{C} k(x | x^{*}) \, dx
\] 
and some $x^{*} \in \mathcal{X}$.  If for each $x'$, $k(\cdot | x')$
is positive and continuous on $C$, then there exists a minorization
condition for $k$.
\end{lemma}

\begin{proof}
  The proof follows a technique first introduced by
  \citet{mykl:tier:yu:1995}.  Fix $x^{*} \in \mathcal{X}$.  Then for
  all $x \in C$
\[
k(x|x') = \frac{k(x|x^{*})}{k(x|x^{*})}k(x|x') \ge \left[ \inf_{x \in
    C} \frac{k(x|x')}{k(x|x^{*})} \right] k(x|x^{*})  \; .
\]
Let $x_{m}$ be the point where the infimum is achieved.  Then the
minorization follows by setting $q(x) = c^{-1} k(x | x^{*}) I(x \in
C)$ and
\[
s(x') = c \frac{k(x_{m}|x')}{k(x_{m}|x^{*})}  \; .
\]
\end{proof}

The conditions of Lemma~\ref{lem:silly minorization} are not the
weakest that ensure the existence of a minorization condition but they
will suffice for our purposes.  In particular, it is straightforward
to use Lemma~\ref{lem:silly minorization} to see that there exists a
minorization condition for both $k_{1}$ and $k_{2}$ the Mtd's for
$\Phi_{1}$ and $\Phi_{2}$, respectively.  Also,
\citet{hobe:jone:robe:2006} derived an explicit closed form expression
for a minorization for a Markov chain for which $\Phi_{2}$ is a
special case.

The consistency results for $\hat{\sigma}^{2}_{n}$ in
\citet{fleg:jone:2009}, \citet{hobe:etal:2002},
\citet{jone:hara:caff:neat:2006} and \citet{bedn:latu:2007} all
require that a minorization condition hold.  The efficacy of
regenerative simulation is utterly dependent upon the minorization
while minorization is irrelevant to the implementation of batch means
and spectral methods.  That is, the minorization is purely a technical
device used in the proofs of consistency for batch means and spectral
estimators.

We use the method of batch means in Section~\ref{sec:example} to
estimate $\sigma^{2}_{g}$.  Let $n$ be the simulation length,
$b_{n}=\lfloor n^{a}\rfloor$ and $a_{n} = \lfloor n / b_{n} \rfloor$.
Now define  
\begin{equation*} 
\bar{Y}_{j} := \frac{1}{b_{n}} \sum_{i=(j-1)b_{n}}^{jb_{n}-1} g
(\xi_{i}, \lambda_{i}) \hspace*{5mm} \text{ for } j=1,\ldots,a_{n} \;
.   
\end{equation*} 
The batch means estimate of $\sigma_{g}^{2}$ is 
\begin{equation} \label{eq:bmvar} \hat{\sigma}_{n}^{2} =
  \frac{b_{n}}{a_{n}-1} \sum_{j=1}^{a_{n}} (\bar{Y}_{j} -
  \bar{g}_n)^{2} \; .   
\end{equation}
Putting together our Theorem~\ref{thm:geo1} and Lemma~\ref{lem:silly
  minorization} with results in \citet{jone:hara:caff:neat:2006} and
\citet{bedn:latu:2007} we have the following consistency result.
\begin{theorem} \label{thm:batch means} Assume the conditions of
  Theorem~\ref{thm:geo1}.  If $E_{\pi} |g|^{2+\epsilon} < \infty$ for
  some $\epsilon > 0$ set $\epsilon=\epsilon_{1} + \epsilon_{2}$ and
  let $(1 + \epsilon_{1} / 2)^{-1} < a < 1$, then for any initial
  distribution for either $\Phi_{1}$ or $\Phi_{2}$ we have that
  $\hat{\sigma}_{n}^{2} \to \sigma_{g}^{2} $ with probability 1 as $n
  \to \infty$.
\end{theorem}

Using Theorems~\ref{thm:clt} and~\ref{thm:batch means} we can use
\eqref{eq:bmvar} to form an asymptotically valid confidence interval
for $E_{\pi}g$ in the usual way 
\begin{equation} \label{eq:ci} 
\bar{g}_{n} \pm t_{a_{n}-1} \frac{\hat{\sigma}_{n}}{\sqrt{n}} 
\end{equation} 
where $t_{a_{n}-1}$ is a quantile from a Student's $t$ distribution
with $a_{n}-1$ degrees of freedom.  Moreover, we can use batch means
to implement the fixed-width methods of
\citet{jone:hara:caff:neat:2006} to determine how long to run the
simulation.  Following \citet{fleg:hara:jone:2008} let $\varepsilon$
be the desired  half-width of the interval in \eqref{eq:ci} and
$n^{*}$ be a minimum simulation size specified by the user.  Then we
can terminate the simulation the first time 
\[
 t_{a_{n}-1} \frac{\hat{\sigma}_{n}}{\sqrt{n}} + \varepsilon I(n \ge
 n^{*}) + \frac{1}{n} \le \varepsilon \; .
\]
The final interval estimate will be asymptotically valid in the sense
that the interval will have the desired coverage probability for
sufficiently small $\varepsilon$; see also \citet{fleg:hara:jone:2008},
\citet{fleg:jone:2009}, \citet{glyn:whit:1992} and
\citet{jone:hara:caff:neat:2006}. 

\section{A Numerical Example}\label{sec:example}

\begin{figure}[h]\centering
\centering
 \includegraphics[height=3.5in,width=4.25in]{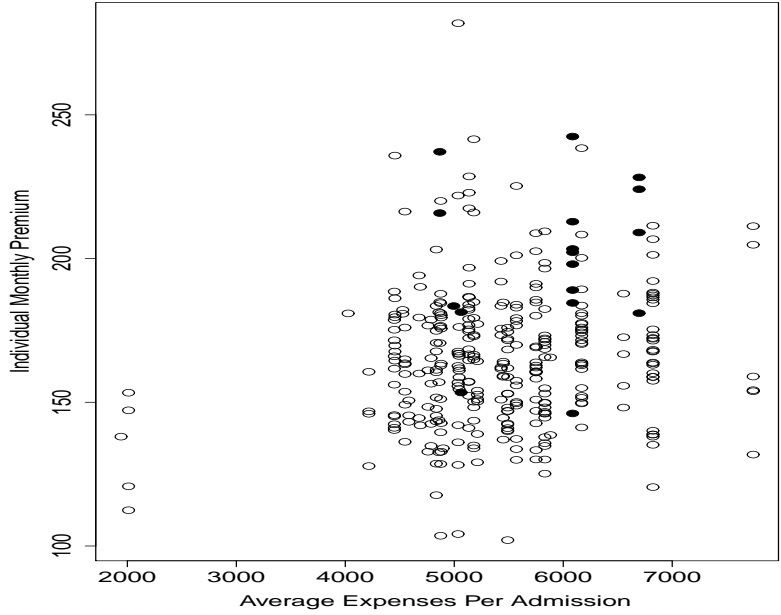}
 \caption{Individual monthly HMO premiums are plotted against the
   average expenses per admission in the state in which the HMO
   operates.  Solid circles represent states in New England.}
 \label{plot:hmoscatter}
\end{figure}

In this section we illustrate our theoretical results in the analysis
of US government health maintenance organization (HMO) data.  To study
the cost-effectiveness of transferring military retirees from a
Defense Department health plan to health plans for government
employees, information was gathered from 341 state-based health
maintenance organizations (HMOs).  These plans represent 42 states,
the District of Columbia, Puerto Rico, and Guam.  An HMO plan's cost
is measured by its monthly premium for individual subscribers.  Two
possible factors in this cost are (1) the typical hospital expenses in
the state in which the HMO operates; and (2) the region in which the
HMO operates.  In Figure \ref{plot:hmoscatter}, the individual monthly
premiums for the 341 HMOs are plotted against the average expenses per
admission in the state of operation (both in US dollars).

Let $y_i$ denote the individual monthly premium of the $i$th HMO plan.
To analyze these data, \citet{hodg:1998} considered a Bayesian version
of the following frequentist model: 
\begin{equation}\label{eq:freq}
y_i = \beta_0 + \beta_1 x_{i1} + \beta_2 x_{i2} + \varepsilon_i
\end{equation}
where the $\varepsilon_i$ are iid
$\text{N}\left(0,\lambda_R^{-1}\right)$, $x_{i1}$ denotes the centered
and scaled average expenses per admission in the state in which the
$i$th HMO operates, and $x_{i2}$ is an indicator for New England.  The
$x_{i1}$ values were centered and scaled to avoid collinearity.
Specifically, if $\tilde{x}_{i1}$ is the raw average expense per
admission and $\overline{x}_{1}$ is the overall average expense per
admission, $x_{i1} = (\tilde{x}_{i1} - \overline{x}_{1})/1000$.

We perform a Bayesian regression analysis based on the following
hierarchical version of \eqref{eq:freq}: 
\begin{equation}\label{eq:hier3}
\begin{split}
 y|\beta,\lambda_R & \sim \text{N}_N\left(X\beta, \lambda_R^{-1}I_N\right) \\
 \beta|\lambda_R   & \sim \text{N}_3\left(b,B^{-1}\right) \\ 
 \lambda_R         & \sim \text{Gamma}(r_1,r_2) \\ 
\end{split} 
\end{equation}
where $N=341$, $y$ is the $N\times 1$ vector of individual premiums,
$\beta=(\beta_0,\beta_1,\beta_2)$ is the vector of regression
parameters, and $X$ is the $N\times 3$ data matrix.

Complete specification of the model requires values for
hyperparameters $(b, B, r_1, r_2)$.  We use an approach which is
empirical Bayesian in spirit.  To this end, we fit \eqref{eq:freq}
using least squares regression.  The results are summarized in Table
\ref{tab:lsr}.  
\begin{table}[h] \caption{Least squares regression results for
    \eqref{eq:freq}.}  \centering
  \renewcommand{\arraystretch}{1.25} \begin{tabular}{crc} \hline 
    Parameter & Estimate & Standard Error \\
    \hline
    $\beta_0$ & 164.989  & 1.322 \\
    $\beta_1$ &   3.910  & 1.508 \\
    $\beta_2$ &  32.799  & 5.961 \\
    \hline
    \multicolumn{3}{l}{$N = 341$} \\
    \multicolumn{3}{l}{degrees of freedom = 338} \\
    \multicolumn{3}{l}{$\text{MSE} = \text{SSE}/338 = \sum_{i=1}^N
      (y_i - \hat{y}_i)^2/338 = 23.79^2$} \\ 
    \hline \end{tabular} \label{tab:lsr} \end{table}

Accordingly, we chose the following prior mean and covariance matrix
for $\beta$: 
\[
b = \left(\begin{array}{r} 164.989 \\ 3.910 \\ 32.799  \end{array} \right)
\hspace{4mm} \text{ and } \hspace{4mm}
B^{-1} = \left(\begin{array}{ccc} 2 & 0 & 0 \\ 0 & 3 & 0 \\ 0 & 0 &
    36 \end{array} \right) \;  
\]
where $b$ is the vector of least squares estimates and the diagonal
elements of $B^{-1}$ are reflective of the corresponding squared
standard errors in Table \ref{tab:lsr}.  Next, we set the prior mean
and variance for $\lambda_R$ to
\[
\begin{split}
\text{E}(\lambda_R) & = \frac{r_1}{r_2} = \frac{1}{\text{MSE}} =
0.00177;  \; \text{ and }\\ 
\text{Var}(\lambda_R) & =\frac{r_1}{r_2^2} = 1 \\
\end{split} 
\]
where MSE is the least squares estimate of $\lambda_R^{-1}$ given in
Table \ref{tab:lsr}.  Solving for $r_1$ and $r_2$ gives $r_1 =
3.122*10^{-6}$ and $r_2 = 0.00177$. 

Since \eqref{eq:hier3} does not contain any random effects, it follows
from Theorem \ref{thm:geo1} that the Gibbs sampler for
$\pi(\beta,\lambda_R|y)$ is geometrically ergodic since
\[
r_{1} > 0 \vee 0.5\left[2-N\right] = 0 \; 
\] 
and for any $\lambda_R \in \mathbb{R}_+$ the function $G(\lambda_R)$
is bounded (recall Proposition~\ref{prop:bound1}) where
\[
G(\lambda_R) = \sum_{i=1}^N \left[\text{E}(y_i - x_i\beta|\lambda_R,y)
\right]^2 = \left(y - X\text{E}(\beta|\lambda_R,y) \right)^T\left(y -
  X\text{E}(\beta|\lambda_R,y) \right)  
\]
and $\text{E}(\beta|\lambda_R,y) = (\lambda_RX^TX +
B)^{-1}(\lambda_RX^Ty + Bb)$. 

Consider estimating the posterior means of $\beta_0$, $\beta_1$, and
$\beta_2$.  By Lemma \ref{lem:exmom}, the fourth posterior moments of
these parameters are finite.  Thus Theorems~\ref{thm:clt}
and~\ref{thm:batch means} in conjunction with geometric ergodicity
guarantee the existence of CLTs and consistent estimators of the
asymptotic variance via batch means with $b_{n}=\lfloor
n^{0.501}\rfloor$ which was chosen based on recommendations in
\citet{jone:hara:caff:neat:2006}.

To begin our analysis of the posterior means, we ran independent Gibbs
samplers from a variety of starting values and updated $\lambda_R$
followed by $\beta$ in each iteration.  For each chain, we required a
minimum simulation length of 1000.  At each successive iteration, we
calculated the approximate half-widths of the Bonferroni-corrected
95\% intervals for the posterior means of $\beta_0$, $\beta_1$, and
$\beta_2$,
\[
 t_{a_{n}-1, \; 0.025/3} \frac{\hat{\sigma}_{n}}{\sqrt{n}}  +
 \frac{1}{n} \; . 
\]
Simulation continued until the half-widths for $\beta_0$, $\beta_1$,
and $\beta_2$, were below 0.10, 0.02, and 0.10, respectively.  The
results were consistent across starting values.  That is, Gibbs
samplers with different starting values produced similar estimates and
required similar simulation effort to meet the above specifications.
Here, we present the results for the chain started from the prior
means of $\beta$ and $\lambda_R$, $(b,r_1/r_2)$.  Under this setting,
the interval half-width thresholds were met after 16831 iterations.
The corresponding estimates of the posterior means are reported in
Table \ref{tab:est1} with standard errors.

\begin{table}[h] \caption{Estimates of posterior means with
    corresponding standard errors.}  \centering
  \renewcommand{\arraystretch}{1.25} \begin{tabular}{|c|r@{}l|c|r@{,}}
    \hline 
    Parameter   &  \multicolumn{2}{|c|}{Estimate}  & Standard Error \\
    \hline
    $\beta_0$   & 165 & .0   & 0.007 \\
    $\beta_1$   & 3 & .9   & 0.008    \\
    $\beta_2$   & 32 & .8   & 0.032   \\    
    \hline \end{tabular} \label{tab:est1} \end{table}

\begin{appendix}
\section{Appendix}

\subsection{Proof of Proposition \ref{prop:drift1}}

We will require the following general results in our proof.  A proof
of Lemma \ref{fact1} is given in \cite{hend:sear:1981} and Lemma
\ref{fact2} follows from the convexity of the inverse function.

\begin{lemma}\label{fact1} 
  Let $A$ be a nonsingular $n \times n$ matrix, $B$ be a nonsingular
  $s \times s$ matrix, $U$ be an $n \times s$ matrix, and $V$ be an $s
  \times n$ matrix.  Then
\[
 (A + UBV)^{-1} = A^{-1} - A^{-1}U(B^{-1} + VA^{-1}U)^{-1}VA^{-1}.
\]
When $U=V$ this implies
\[
 x^T(A + UBV)^{-1}x \le x^TA^{-1}x
\]
for any $n \times 1$ vector $x$.
\end{lemma}

\begin{lemma}\label{fact2}  Let $x$ be an $m \times 1$ vector.  Also,
  let $A$ and $B$ be nonsingular, $m \times m$ matrices.  Then
\[
 x^T(A + B)^{-1}x \le \frac{1}{4}x^T\left(A^{-1} + B^{-1}\right)x \; .
\]   
\end{lemma}

We begin the proof of Proposition \ref{prop:drift1}.  Recall that
\[
 v_1(\xi) := (y-X\beta-Zu)^T(y-X\beta-Zu), \hspace{4mm} \text{ and }
 \hspace{4mm} v_2(\xi) := u^Tu\; .
\]
We must show that for all $\xi' \in \mathbb{R}^{k+p}$
\[
\text{E}_{k_\xi}[V(\xi)|\xi'] = \text{E}_{k_\xi}[v_1(\xi) +
v_2(\xi)|\xi'] \le \gamma V(\xi') + L 
\]
where the constants $\gamma$ and $L$ are given in the statement of
Proposition \ref{prop:drift1}.  Let $\text{E}_i$ and $\text{Var}_i$
denote expectation and variance with respect to the
$N_{k+p}\left(m_{i},\Sigma^{-1}\right)$ distribution.  Similarly, let
$\text{E}_{jl}$ and $\text{Var}_{jl}$ denote expectation and variance
with respect to density
\[
f_{1j}(\lambda_R|\xi,y) f_{2l}(\lambda_D|\xi,y)
\] 
defined by \eqref{eq:FullCond1}.
Notice that 
\begin{equation}\label{eq8}
 \text{E}_{k_\xi}\left[V(\xi)|\xi'\right] 
 =  \text{E}\left[\text{E}(V(\xi)|\lambda)|\xi'\right]  
 = \sum_{j=1}^s\sum_{l=1}^t \sum_{i=1}^r \phi_j \psi_l \eta_i
 \text{E}_{jl}\left[ \text{E}_i (V(\xi)|\lambda) \; \vline \;
   \xi'\right]  
\end{equation}
where the first equality holds by the construction of $\Phi_\xi$.
Thus we focus on the $\text{E}_{jl}\left[ \text{E}_i (V(\xi)|\lambda)
  \; \vline \; \xi'\right]$ in the next 3 lemmas.

\begin{lemma} \label{lemma:v1drift}
Suppose $Z^TZ$ is nonsingular.  Then for all $i,j,l$
\[
  \text{E}_{jl}\left[ \text{E}_i (v_{1}(\xi)|\lambda) \; \vline \; \xi'\right] \le  \delta_{j1} v_1(\xi')  + L_1  
\]
where
\begin{equation}\label{eq:b1}
 L_1 = \text{E}_{jl}\left[ \sum_{m=1}^N 
   [\text{E}_i(y_m-x_m\beta-z_mu|\lambda)]^2 \;\vline\;
   \xi'\right]  + \sum_{m=1}^N x_mB^{-1}x_m^T + 2r_{j2}\delta_{j1} . 
\end{equation}
\end{lemma}

\begin{proof}
Consider the inner expectation $\text{E}_i (v_{1}(\xi)|\lambda)$.  For any $i$ we have 
\[
\begin{split}
\text{E}_i (v_{1}(\xi)|\lambda)
 & =   \sum_{m=1}^N \text{E}_i\left[(y_m-x_m\beta-z_mu)^2|\lambda\right] \\
 & =   \sum_{m=1}^N \left[\text{E}_i(y_m-x_m\beta-z_mu|\lambda)\right]^2 + \sum_{m=1}^N \text{Var}_i(y_m-x_m\beta-z_mu|\lambda)\\
\end{split}
\]
and 
\[
\begin{split}
   \text{Var}_i(y_m-x_m\beta-z_mu|\lambda)
     & = x_m\left(\lambda_RX^TX + B\right)^{-1}x_m^T \\
     & \hspace{6mm} +    z_m\left(\lambda_RZ^TZ + \lambda_DI_k\right)^{-1}z_m^T   \\
     & \le x_mB^{-1}x_m^T +\lambda_R^{-1}z_m\left(Z^TZ\right)^{-1}z_m^T \\
 \end{split}
\]
by Lemma~\ref{fact1}.  It follows that for any $i,j,l$ we have 
\[
\begin{split}
\text{E}_{jl}\left[ \text{E}_i (v_{1}(\xi)|\lambda) \; \vline \; \xi'\right] 
 & \le \text{E}_{jl}\left[  \sum_{m=1}^N\left[\text{E}_i(y_m-x_m\beta-z_mu|\lambda)\right]^2\; \vline \; \xi'\right] \\
 & \hspace{6mm} + \text{E}_{jl}\left(\lambda_R^{-1} | \xi'\right)\sum_{m=1}^N z_m\left(Z^TZ\right)^{-1}z_m^T + \sum_{m=1}^N x_mB^{-1}x_m^T  \; .\\
\end{split}
\]
Combining this with the fact that 
\[
 \text{E}_{jl}\left(\lambda_R^{-1} | \xi'\right) 
   = \frac{2r_{j2} + v_1(\xi')}{2r_{j1} + N - 2}  
   = \frac{\delta_{j1}(2r_{j2} + v_1(\xi'))}{\sum_{m=1}^N z_m\left(Z^TZ\right)^{-1}z_m^T}\;
\]
gives
\[
\begin{split}
\text{E}_{jl}\left[ \text{E}_i (v_{1}(\xi)|\lambda) \; \vline \; \xi'\right] 
 & \le  \text{E}_{jl}\left[ \sum_{m=1}^N\left[\text{E}_i(y_m-x_m\beta-z_mu|\lambda)\right]^2\; \vline \; \xi'\right] \\
 & \hspace{6mm} + \delta_{j1}(2r_{j2} + v_1(\xi')) + \sum_{m=1}^N x_mB^{-1}x_m^T  \\
 & = \delta_{j1} v_1(\xi')  + L_1  \; . \\
\end{split}
\]
\end{proof}

\begin{lemma} \label{lemma:v1v2drift}
For any $i,j,l$
\[
 \text{E}_{jl}\left[ \text{E}_i (v_{1}(\xi)|\lambda) \; \vline \; \xi'\right]  \le \delta_{j3}v_1(\xi') +
 (\delta_{l4}-\delta_{l2})v_2(\xi') + L_2
\]
where
\begin{equation}\label{eq:b3}
  L_2 =  \text{E}_{jl}\left[ \sum_{m=1}^N
   [\text{E}_i(y_m-x_m\beta-z_mu|\lambda)]^2 \;\vline\;
   \xi'\right] + \frac{1}{4}\sum_{m=1}^N x_mB^{-1}x_m^T +
 2r_{j2}\delta_{j3} + 2d_{l2}(\delta_{l4}-\delta_{l2}).  
\end{equation}
\end{lemma}

\begin{proof}
Notice that for any $i,j,l$
\begin{equation}\label{eq3}
\begin{split}
\text{E}_{jl}\left[ \text{E}_i (v_{1}(\xi)|\lambda) \; \vline \; \xi'\right] 
& =   \text{E}_{jl}\left[  \sum_{m=1}^N\text{E}_i\left[\left(y_m-x_m\beta-z_mu\right)^2|\lambda\right] \; \vline \; \xi'\right] \\
& =   \text{E}_{jl}\left[  \sum_{m=1}^N\left[\text{E}_i(y_m-x_m\beta-z_mu|\lambda)\right]^2 \; \vline \; \xi'\right] \\
& \hspace{6mm} + \sum_{m=1}^N \text{E}_{jl}\left[ \text{Var}_i(y_m-x_m\beta-z_mu|\lambda) \; \vline \; \xi'\right] \\
\end{split}
\end{equation}
where from Lemmas~\ref{fact1} and~\ref{fact2} 
\[
\begin{split}
  \text{Var}_i(y_m-x_m\beta-z_mu|\lambda) & = x_m\left(\lambda_RX^TX +
    B\right)^{-1}x_m^T + z_m\left(\lambda_RZ^TZ +
    \lambda_DI_k\right)^{-1}z_m^T \\
 & \le \frac{1}{4}x_m\left(\lambda_R^{-1}\left(X^TX\right)^{-1} +
    B^{-1}\right)x_m^T + \lambda_D^{-1}z_mz_m^T \; .
\end{split}
\]
Also, by \eqref{eq:FullCond1} we have
\[
 \text{E}_{jl}\left(\lambda_R^{-1} | \xi'\right) = \frac{2r_{j2} +
   v_1(\xi')}{2r_{j1} + N - 2} \hspace{2mm} \text{  and  } \hspace{2mm}
 \text{E}_{jl}\left(\lambda_D^{-1} | \xi'\right) = \frac{2d_{l2} +
   v_2(\xi')}{2d_{l1} + k - 2}. 
\]
Therefore

\vspace{3mm}
\noindent $\sum_{m=1}^N \text{E}_{jl}\left[ \text{Var}_i(y_m-x_m\beta-z_mu|\lambda) \;\vline\;
    \xi'\right]$

\vspace{-3mm}
\begin{equation}\label{eq4}
\begin{split}
   & \le \sum_{m=1}^N \left[\frac{1}{4}x_m\left(\frac{2r_{j2} + v_1(\xi')}{2r_{j1}
      + N - 2}\left(X^TX\right)^{-1} + B^{-1}
  \right)x_m^T  + \frac{2d_{l2} + v_2(\xi')}{2d_{l1} + k -
    2}z_mz_m^T \right]\; \\
   & = \delta_{j3}\left(2r_{j2} + v_1(\xi')  \right) +
  \frac{1}{4}\sum_{m=1}^N x_m B^{-1}x_m^T  + \left(2d_{l2} + v_2(\xi')\right) \frac{\sum_{m=1}^N
    z_mz_m^T}{2d_{l1} + k - 2}  \\
   & = \delta_{j3}\left(2r_{j2} + v_1(\xi')  \right) +
  \frac{1}{4}\sum_{m=1}^N x_m B^{-1}x_m^T   + \left(2d_{l2} + v_2(\xi')\right) (\delta_{l4}-\delta_{l2}) \; . \\
 \end{split}
\end{equation}
The result holds by combining \eqref{eq3} and \eqref{eq4}.
\end{proof}

\begin{lemma} \label{lemma:v2drift}
For any $i,j,l$
\[
 \text{E}_{jl}\left[ \text{E}_i (v_{2}(\xi)|\lambda) \; \vline \; \xi'\right] \le  \delta_{l2} v_2(\xi') + L_3
\]
where
\begin{equation}\label{eq:b2}
 L_3 =   \text{E}_{jl}\left[\sum_{m=1}^k
   [\text{E}_i\;(u_m|\lambda)]^2 \;\vline\; \xi'\right] + 2d_{l2}\delta_{l2}. 
\end{equation}
\end{lemma}

\begin{proof}
First, for any $i,j,l$
\begin{equation}\label{eq5}
\begin{split}
  \text{E}_{jl}\left[ \text{E}_i (v_{2}(\xi)|\lambda) \; \vline \; \xi'\right]  
   & =    \text{E}_{jl}\left[ \sum_{m=1}^k\text{E}_i\left(u_m^2|\lambda\right) \;\vline\;    \xi'\right] \\
   & = \text{E}_{jl}\left[\sum_{m=1}^k    [\text{E}_i\;(u_m|\lambda)]^2 \;\vline\; \xi'\right]  + \sum_{m=1}^k \text{E}_{jl}\left[ \text{Var}_i(u_m|\lambda)\;\vline\;     \xi'\right] \; . \\
\end{split}
\end{equation}
Let $e_m$ denote the $k \times 1$ vector with the $m$th element being
1 and the rest of the elements being 0.  Thus by
Lemma~\ref{fact1}, 
\begin{equation}\label{eq6}
 \text{Var}_i(u_m|\lambda) =   e_m^T\left(\lambda_RZ^TZ + \lambda_DI_k
 \right)^{-1}e_m \; \le \; \lambda_D^{-1} e_m^Te_m  \; = \; \lambda_D^{-1} \; .
\end{equation}
Also, 
\begin{equation}\label{eq7}
 \text{E}_{jl}\left(\lambda_D^{-1} | \xi' \right) = \frac{2d_{l2} +
   v_2(\xi')}{2d_{l1} + k - 2} = \frac{\delta_{l2}}{k} \left(2d_{l2} +
   v_2(\xi')\right). 
\end{equation}
Putting \eqref{eq5}--\eqref{eq7} together gives
\[
 \begin{split}
 \text{E}_{jl}\left[ \text{E}_i (v_{2}(\xi)|\lambda) \; \vline \; \xi'\right] 
  & \le   \text{E}_{jl}\left[ \sum_{m=1}^k  [\text{E}_i\;(u_m|\lambda)]^2 \;\vline\; \xi'\right]  + \sum_{m=1}^k \text{E}_{jl}\left[ \lambda_D^{-1}\;\vline\;     \xi'\right] \\
  &=  \text{E}_{jl}\left[  \sum_{m=1}^k [\text{E}_i\;(u_m|\lambda)]^2 \;\vline\; \xi'\right]  + \delta_{l2}\left(2d_{l2}+v_2(\xi')\right) \\ 
   &= \delta_{l2}v_2(\xi') + L_3 \; . \\
    \end{split}
\]
\end{proof}

We are now ready to finish the proof of Proposition \ref{prop:drift1}.
We consider the case with nonsingular $Z^TZ$ and the case in which no
restrictions are placed on $Z$ separately.

\begin{enumerate}
\item Case 1: $Z^TZ$ nonsingular

  Notice that $L_1 + L_3 \le L$ for $L_1$ and $L_3$ given by
  \eqref{eq:b1} and \eqref{eq:b2}, respectively.  Then by Lemmas
  \ref{lemma:v1drift} and \ref{lemma:v2drift} we have that for any
  $i,j,l$
\begin{equation}\label{eq9}
\begin{split}
\text{E}_{jl}\left[ \text{E}_i (V(\xi)|\lambda) \; \vline \; \xi'\right] 
 & =   \text{E}_{jl}\left[ \text{E}_i (v_1(\xi)+v_2(\xi)|\lambda) \;
   \vline \; \xi'\right] \\ 
 & \le \delta_{j1} v_1(\xi') + \delta_{l2} v_2(\xi') + L_1 + L_3 \\
 & \le \gamma V(\xi') + L \; .
\end{split}
\end{equation}
Then combining \eqref{eq8} and \eqref{eq9} establishes the drift condition.

\item Case 2: $Z^TZ$ is possibly singular

  Observe that $L_2 + L_3 \le L$ for $L_2$ and $L_3$ given by
  \eqref{eq:b3} and \eqref{eq:b2}, respectively.  Further, it follows
  from Lemmas \ref{lemma:v1v2drift} and \ref{lemma:v2drift} that for
  any $i,j,l$
\begin{equation}\label{eq11}
\begin{split}
\text{E}_{jl}\left[ \text{E}_i (V(\xi)|\lambda) \; \vline \; \xi'\right] 
 & =   \text{E}_{jl}\left[ \text{E}_i (v_1(\xi)+v_2(\xi)|\lambda) \;
   \vline \; \xi'\right] \\ 
 & \le \delta_{j3} v_1(\xi') + (\delta_{l4}- \delta_{l2}) v_2(\xi') +
 L_2 + \delta_{l2}v_2(\xi') + L_3 \\ 
 & = \delta_{j3} v_1(\xi') + \delta_{l4}v_2(\xi') + L_2 + L_3\\
 & \le \gamma V(\xi') + L \; .
\end{split}
\end{equation}
Hence the result holds by combining \eqref{eq8} and \eqref{eq11}.
\end{enumerate}

\subsection{Proof of Proposition~\ref{prop:bound1}}
\label{app:bound}

By the assumption that $b_i=0$ for all $i$, $\xi|\lambda,y \sim
\text{N}_{k+p}(m_{0},\Sigma^{-1})$ 
where \[
\begin{split}
\Sigma^{-1} & = \left( \begin{array}{cc}
\left(\lambda_RZ^TZ + \lambda_DI_k\right)^{-1} & 0 \\
0               & \left(\lambda_RX^TX+B\right)^{-1} \\
\end{array} \right) \\
m_{0} & = \left( \begin{array}{c}
\lambda_R\left(\lambda_RZ^TZ + \lambda_DI_k\right)^{-1}Z^Ty \\
\lambda_R \left(\lambda_RX^TX + B \right)^{-1}X^Ty \\ 
\end{array} \right) \; . \\
\end{split}
\]
Define $A_g := \lambda_RX^TX+B$ and $A_h := \lambda_RZ^TZ +
\lambda_DI_k$.  
Then $\text{E}(\beta|\lambda) = \lambda_RA_g^{-1}X^Ty$ and
$\text{E}(u|\lambda) = \lambda_R A_h^{-1}Z^Ty$. 

We must establish that there exists $K$ for which
\[
    \sum_{m=1}^N\left[ \text{E}\;(y_m-x_m\beta-z_mu|\lambda,y)\right]^2 + 
  \sum_{m=1}^k\left[ \text{E}\;(u_m|\lambda,y)\right]^2  \le K \; .
\]
Let
\[
f(\lambda) = (y - X \text{E}(\beta | \lambda) - Z \text{E}(u |
\lambda))^{T} (y - X \text{E}(\beta | \lambda) - Z \text{E}(u |
\lambda)) + \text{E}(u | \lambda)^{T}\text{E}(u | \lambda) 
\]
and note that the claim will be proven if we can show that $f(\lambda)
\le K$ for all $\lambda$.  To this end, define functions $g$,
and $h$ as
\[
 \begin{split}
   g(\lambda) & =
   (y - X \text{E}(\beta|\lambda))^T (y - X \text{E}(\beta|\lambda)) \\
   h(\lambda) & = \text{E}(u|\lambda)^TZ^TZ\text{E}(u|\lambda) +
   \text{E}(u|\lambda)^T\text{E}(u|\lambda) -
   2y^T Z \text{E}(u|\lambda) \; .
 \end{split}
\]
Since the conditional independence of $\beta$ and $u$ given $\lambda$
implies $X^TZ=0$, a little algebra shows that $f(\lambda) = g(\lambda)
+ h(\lambda)$.  Thus, it suffices to find $K_g$ and $K_h$
such that for all $\lambda$, $g(\lambda) \le K_g^2$ and
$h(\lambda) \le K_h^2$.

First,
\[
 \begin{split}
g(\lambda)  &= y^Ty + \text{E}( \beta | \lambda)^T X^T X
\text{E}(\beta | \lambda) - 2y^T X \text{E}(\beta|\lambda) \\ 
  &= y^Ty + \lambda_R^2y^TXA_g^{-1}X^TXA_g^{-1}X^Ty -
  2\lambda_R y^T X A_g^{-1} X^T y \\ 
  &= y^Ty - \lambda_Ry^TXA_g^{-1}BA_g^{-1}X^Ty + 
  \lambda_Ry^TXA_g^{-1}A_gA_g^{-1}X^Ty \\ 
  &\hspace{4mm}- 2\lambda_Ry^TXA_g^{-1}X^Ty \\
  &= y^Ty - \lambda_Ry^TXA_g^{-1}BA_g^{-1}X^Ty -
  \lambda_Ry^TXA_g^{-1}X^Ty \\ 
  & \le y^Ty \\
& := K_{g}^{2}
 \end{split}
\]
by the positive definiteness of $B$ and $A_g^{-1}.$ Notice that this
also proves part 1 of the claim.  Next, we have
\[
 \begin{split}
   h(\lambda)& = \lambda_R^2 y^TZA_h^{-1} Z^TZ A_h^{-1} Z^Ty +
   \lambda_R^2 y^TZ A_h^{-2}Z^Ty - 2\lambda_R y^TZ A_h^{-1}Z^Ty \\ 
   & = \lambda_R y^TZA_h^{-1}A_hA_h^{-1}Z^Ty - \lambda_R\lambda_D
   y^TZA_h^{-2}Z^Ty + \lambda_R^2y^TZA_h^{-2}Z^Ty \\ 
   & \hspace{4mm}- 2\lambda_Ry^TZA_h^{-1}Z^Ty \\
   & = (\lambda_R^2 - \lambda_R\lambda_D) y^TZA_h^{-2}Z^Ty  -
   \lambda_Ry^TZA_h^{-1}Z^Ty. \\ 
\end{split}
\]
Since $A_h^{-1}$ and $A_h^{-2}$ are positive semidefinite we have 
\[
 \begin{split}
h(\lambda)
  & \le \lambda_R^2 y^TZA_h^{-2}Z^Ty \\
  & = \lambda_R^2 y^TZ\left(\left(\lambda_RZ^TZ\right)^2 +
    \lambda_D\left(2\lambda_RZ^TZ +
      \lambda_DI_k\right)\right)^{-1}Z^Ty \\ 
  & \le \lambda_R^2 y^TZ(\lambda_RZ^TZ)^{-2}Z^Ty \\
  & = y^TZ(Z^TZ)^{-2}Z^T y \\
& := K_{h}^{2}
 \end{split}
\]
where the last inequality holds by Lemma~\ref{fact1}. The result now
follows by setting $K^{2} = K_{g}^{2} + K_{h}^{2}$.

\subsection{Lemma \ref{lem:exmom}}
\begin{lemma}\label{lem:exmom}
The fourth posterior moments of $\beta_0$, $\beta_1$, and $\beta_2$ are each finite.
\end{lemma}
\begin{proof}
We present the proof for $\beta_2$.  The proofs for $\beta_0$ and $\beta_1$ are similar.
The finiteness of $\text{E}\left[\beta_2^4 \; \vline \; y \right]$ will follow from establishing that $\text{E}\left[ \beta_2^4 \; \vline \; \lambda_R, y \right]$ is finite since
\[
\text{E}\left[ \beta_2^4 \; \vline \; y \right] = \text{E}\left[ \text{E}\left( \beta_2^4 \; \vline \; \lambda_R, y\right) \; \vline \; y \right] \; . 
\]
To this end, recall that
\[
\beta|\lambda_R,y \sim N\left(\left(\lambda_R X^TX+B\right)^{-1}\left(\lambda_R X^Ty + Bb\right), \; \left(\lambda_R X^TX + B\right)^{-1} \right) \; .
\]
Also, let $\mu_2= \text{E}(\beta_2|\lambda_R,y)$ and $e_3$ denote a vector of zeroes with a one in the third position.
Then
\[
\begin{split}
\text{E}\left[ \left(\beta_2 - \mu_2\right)^4 \; \vline \; \lambda_R, y\right]
& = 3 \left[\left(\lambda_R X^TX + B \right)^{-1}_{33} \right]^2 \\
& = 3 \left[e_3^T\left(\lambda_R X^TX + B \right)^{-1}e_3 \right]^2 \\
& \le 3 \left[e_3^TB^{-1}e_3 \right]^2 \\ 
& = 3 B^{-2}_{33} \\ 
\end{split}
\]
where the inequality follows from Lemma \ref{fact1}.  It follows that the fourth (non-central) moment $\text{E}\left[ \beta_2^4 \; \vline \; \lambda_R, y\right]$ is finite.
\end{proof}

\end{appendix}

\bibliographystyle{ims}
\bibliography{myref}

\end{document}